# Kerfuffle: a web tool for multi-species gene colocalization analysis

Robert Aboukhalil[1,2], Bernard Fendler[1,3] and Gurinder S. Atwal[1,4]
[1]Cold Spring Harbor Laboratory, Cold Spring Harbor, NY 11724, USA
[2]raboukha@cshl.edu, corresponding author
[3]bfendler@cshl.edu
[4]atwal@cshl.edu

**Abstract**

*Background:*

The evolutionary pressures that underlie the large-scale functional organization of the genome are not well understood in eukaryotes. Recent evidence suggests that functionally similar genes may colocalize (cluster) in the eukaryotic genome, suggesting the role of chromatin-level gene regulation in shaping the physical distribution of coordinated genes. However, few of the bioinformatic tools currently available allow for a systematic study of gene colocalization across several, evolutionarily distant species. Furthermore, most tools require the user to input manually curated lists of gene position information, DNA sequence or gene homology relations between species. With the growing number of sequenced genomes, there is a need to provide new comparative genomics tools that can address the analysis of multi-species gene colocalization.

*Results:*

Kerfuffle is a web tool designed to help discover, visualize, and quantify the physical organization of genomes by identifying significant gene colocalization and conservation across the assembled genomes of available species (currently up to 47, from humans to worms). Kerfuffle only requires the user to specify a list of human genes and the names of other species of interest. Without further input from the user, the software queries the *e!*Ensembl BioMart server to obtain positional information and discovers homology relations in all genes and species specified. Using this information, Kerfuffle performs a multi-species clustering analysis, presents downloadable lists of clustered genes, performs Monte Carlo statistical significance calculations, estimates how conserved gene clusters are across species, plots histograms and interactive graphs, allows users to save their queries, and generates a downloadable visualization of the clusters using the Circos software. These analyses may be used to further explore the functional roles of gene clusters by interrogating the enriched molecular pathways associated with each cluster.

*Conclusions:*

Kerfuffle is a new, easy-to-use and publicly available tool to aid our understanding of functional genomics and comparative genomics. This software allows for flexibility and quick investigations of a user-defined set of genes, and the results may be saved online for further analysis. Kerfuffle is freely available at http://atwallab.org/kerfuffle, is implemented in JavaScript (using jQuery and jsCharts libraries) and PHP 5.2, runs on an Apache server, and stores data in flat files and an SQLite database.

**Background**

Advances in genomics and DNA sequencing technology have fueled growing interest in the large-scale physical and functional organization of chromosomes. Several studies have shown that genomes of many disparate species may have chromosome regions containing clusters of functionally related genes [1-3]. It is well known that operons, ubiquitous in prokaryotes, allow multiple genes to be transcribed at once into a polycistronic mRNA. The extent to which genes colocalize in eukaryotes and the extent to which gene clusters are conserved across species are largely unknown. In eukaryotes, operons are rare [4]; however, there is evidence to suggest that genes within the same biological pathway may be clustered more so than expected by random rearrangements, possibly because of co-regulation [5]. For example, the *Hox* genes are tandem duplicate genes organized into clusters, playing a pivotal role in defining the body plan of organisms. Further, the order of the genes within a *Hox* cluster defines the sequence in which these genes are expressed [6]. While these examples rely on positional clustering, other mechanisms may also lead to gene clusters, for example, clustered genes could be coregulated because (1) their promoters are bound to by the same transcription factors; (2) they share regulatory elements (e.g. bidirectional promoters); and (3) the transcription of a gene can change local chromatin accessibility for its neighbors.

Between evolutionary distinct species, we expect to find random genomic rearrangements that do not conserve gene clusters, unless colocalization is beneficial to the organism. It is possible that colocalization is acted upon by natural selection, conserving the gene clusters across large evolutionary time scales, although it remains unclear what structural, regulatory, and functional factors are responsible for the colocalization [1, 7, 8]. A recent study found that the genome of a number of different species was arranged into neighborhoods of functionally-related genes that were not necessarily orthologous [9].

If functionally related genes cluster for mechanistic purposes, then it is expected that those clustered genes would colocalize in other species as well. How to quantify this conservation is a difficult task since orthologous genes have obtained random changes since their last respective speciation event, are often called different names in other species, often have different or increased/decreased functional attributes and not all genes may be required in other species. To address these related research questions, there is a pressing need for computational tools that can overcome the onerous task of querying the growing list of available assembled genomes, analyzing the spatial ordering of genes across many species to identify whether they form clusters, and assessing the conservation of these clusters across other species.

One such clustering tool is C-Hunter, a command-line program that clusters genes by genome position and GO categories [10]. While C-Hunter is capable of identifying clusters of genes within a species, it does not incorporate an analysis of conserved clustering across multiple species, and is not intended as a tool to query a general set of genes that don't share GO terms. Other tools, such as CGCV, allow for clustering across many species but require the user to input DNA sequences instead of gene names [11]; subsequently, the web tool performs BLAST searches to find orthologous genes, which adds significant overhead to run-time.

There are related tools which identify regions of synteny, such as EnsemblCompara [12], i-ADHoRe [13], MCScanX [14], Cinteny [15], OrthoClusterDB [16] and Syntenator [17]. These tools are useful for identifying homologous genomic regions between species, but do not include an automated approach for evaluating gene clustering and its conservation across species. Utilizing the software that provides a web API and pre-computed homology results, we chose to use EnsemblCompara in Kerfuffle.

To the best of our knowledge, there are currently no tools available for efficiently verifying whether a given list of genes from one species forms clusters and whether these clusters are conserved across other species. To this end, we have developed and implemented a publicly-available web-tool, Kerfuffle, that efficiently computes various summary statistics of gene clustering across most genomes in the *e!*Ensembl database [18], compares significance of clustering with shuffled null models, and graphically displays the results. The main advantage of Kerfuffle is that it only requires a user to specify human gene names and species of interest. In addition, orthologous gene searches are automated utilizing pre-computed homology from *e!*Ensembl servers, a relative statistic is used to quantify cluster conservation, and the online program permits server-side saving of results for each registered user for later analysis. Furthermore, Kerfuffle can generate a visualization of the clusters using the Circos software [19]. This comprehensive platform is an important step in furthering our understanding of genome organization and its evolution.

**Implementation**

Because Kerfuffle is available as a web application, this obviates the need for compiling or installing, is accessible from anywhere, is supported by most web browsers (tested on Firefox, Chrome, Internet Explorer and Safari), and allows us to improve our software on our end without requiring the user to download an update.

The back-end runs on PHP 5.2 on an Apache server. The front-end was built primarily in HTML and JavaScript and two Javascript libraries to enhance user experience, namely jsCharts for plotting graphs and jQuery for Ajax effects. With the list of genes input by the user, Kerfuffle will query the *e!*Ensembl BioMart database and retrieve gene name, ID, chromosomal position and homology information of all genes selected and for each species selected. To improve speed, all the species are queried at once in a parallel fashion, and the results are displayed on screen as soon as they are processed by Kerfuffle.

With the gene positions, Kerfuffle will group the genes into clusters based on their colocality using a clustering algorithm written in C++ to improve performance. Once done, Kerfuffle displays the set of gene clusters and uses the jsCharts library to graph relevant plots and histograms (Figure 2). The plot of Figure 2A is interactive: hovering on each point reveals its x and y coordinate, and clicking on the point will reveal all the gene pairs that are separated by a distance x. Furthermore, p-value calculations are done to estimate statistical significance. To ensure that small p-value requests does not slow down webpage usability, the calculations are performed in the background and once done, the results appear in a table.

If the user inputs the names of human genes and wishes to do an analysis on the genomes of several other species, Kerfuffle will find the corresponding homologs in each of those genomes on its own, with no further user input. Kerfuffle is also flexible in the way it accepts user input. The user may choose to input genes in a textbox one by one or alternatively, may upload a file that contains a list of genes, each of which is separated by a break line. However, we recognize that it is difficult for users to keep track of the dozen file formats they use. Thus, if the uploaded file is a comma- or tab-delimited file with multiple columns, Kerfuffle will ask the user to specify the column in which the gene names are found.

To aid in recurring analyses, we recommend that users create a free Kerfuffle account, in which their results and the queried genes will be saved in our databases. On the back-end, the query results obtained from *e!*Ensembl are temporarily stored in text files and purged every week, unless users decide to save their results to their account, in which case the results remain on the server until the users delete them.

**Results and Discussion**

Kerfuffle provides an option to input a known list of functionally related synapse genes, totaling 477, for demonstrative purposes. The source of these genes may be found in the AmiGO database [20]. Colocalization of these genes is supported by a recent publication that demonstrated clustering of genes associated with GABAergic circuit assembly in the cerebellar cortex of young mice [21]. All analyses and images generated were performed on these genes. The list of genes is available on the Kerfuffle webpage for analysis, located under the "Upload gene list" button labeled "Example: Synapse genes."

**Multi-species Colocalization Analyses**

Kerfuffle allows the user to specify a list of gene names and select up to, currently, 47 species for which the analysis will be performed. The gene identifiers supported by Kerfuffle are *e!*Ensembl gene names or WikiGene names and the orthologs of the input human genes are obtained from EnsemblCompara, which uses maximum likelihood phylogenetic trees for homology prediction. Default analysis parameters are provided, although customization is allowed; parameters include: **(1)** *d,* the maximum number of total intervening genes (or gaps) allowed in a cluster (Figure 1); **(2)** the maximum value on the *x*-axis of the histogram of distance between consecutive gene pairs (Figure 2A, 2C); **(3)** the maximum value on the *x*-axis of the histogram of cluster sizes (Figure 2B); and **(4)** the target size of the minimum p-value.

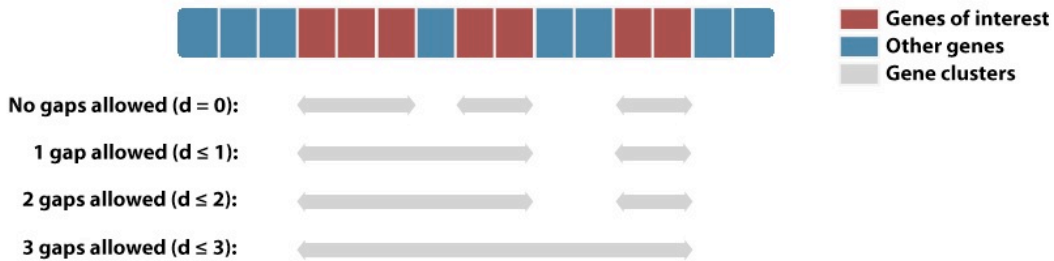

**Figure 1:** Examples of cluster definition. The clusters are defined by the parameter *d*, the maximum number of allowed gaps in a sequence of genes. Red boxes represent queried genes and blue boxes represents genes not queried.

Once the analysis is launched, the local machine will concurrently send asynchronous requests to our web server, one request for each species. For each request, our web server will connect to the *e!*Ensembl BioMart database and download necessary genomic information, after which a colocalization analysis is performed.

Once the analysis is completed, the results are displayed in the "Consecutive gene pairs" (Figure 2A) and "Cluster size distribution" (Figure 2B) graphs. We define the consecutive gene pair distance distribution (Figure 2A) as the histogram of distance between consecutive genes, for all the genes input. The null-distribution in Figure 2A is determined through the following procedure. First, we randomly distribute the genes uploaded to Kerfuffle across the genome. Second, the distance distribution is found for those genes. Finally, this process is iterated many times, dependent on the user's minimum target p-value, and then all random distance distributions are averaged. This average approximates the null-model—that the list of genes do not cluster more so than random. The p-values are determined through a similar process. Upon each random iteration, the counts for each distance in the distribution from the real genomic positions (from uploaded genes) and from the randomly distributed genes are compared. The p-values are calculated as the frequency that random permutation counts surpass real data counts.

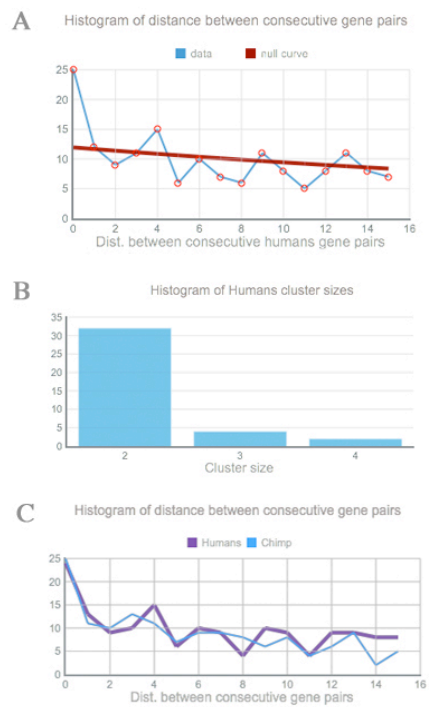

**Figure 2:** Sample output using genes with ontology term 'synapse'. **A.** The discovered (light) and expected (dark) human distance distribution. **B.** Cluster size histogram for humans. **C.** Distance distribution of two species, human (dark) and chimp (light).

Below the p-value table, the cluster size distribution plot (Figure 2B) is displayed with an option to download the data in a text file. Clusters used in the cluster size distribution plot and the conservation analysis below are defined as in Figure 1. Namely, a set of ordered genes $G_1, G_2, ..., G_n$ is said to colocalize, i.e. form a cluster, if the number of total intervening genes is less than or equal to the specified parameter $d$. Mathematically, if $x(G_i)$ is the positional order of gene $G_i$, then we require that $x(G_n) - x(G_1) - n + 1 \leq d$. In Kerfuffle, the default value of the parameter $d$ is 2.

**Comparative Analysis**

In the "Summary" tab, the user can launch an analysis comparing human clusters to those of other species, as well as plot distance histograms between consecutive gene pairs for all species selected (Figure 2C). To find the orthologs of the input human genes, Kerfuffle fetches data from *e!*Ensembl's EnsemblCompara resource [12]. Once selected, the results can be found under the "Compare" tab. Further, to quantify cluster conservation between chosen species, we define a "conservation score" which conveys similarity of clusters among species.

To quantify the conservation of gene clusters in species $T$ relative to those found in species $S$, we use the following conservation score:

$$Score(S,T) = \frac{1}{N_S} \sum_i^{N_S} \sum_j^{N_T} \frac{|S_i \cap T_j|}{|S_i|},$$

where $S_i$ and $T_i$ refers to the set of genes in cluster $i$ in species $S$ and $T$, respectively. $N_X$ refers to the total number of clusters in species $X$. All clusters were chosen as size 2 or larger. The intersection between $S_i$ and $T_j$ is defined as the set of common genes between cluster $i$ in species $S$ and cluster $j$ in species $T$. We normalize the size of the intersection by the size of the cluster $S_i$, hence calculating the score relative to species $S$. The inner sum increases if the genes found in cluster *j* of species T are also found in cluster *i* of species *S*, while the outer sum averages those scores over each cluster *i* in species $S$. Thus, *Score(S, T)* is a statistic which increases as the same clusters are observed and remain intact amongst the species investigated in $T$ relative to $S$. Our default setting for this analysis sets *S = Human*.

**Using the Software**

Before the analysis is started, the user must select the species of the gene names used, upload, type, or paste the genes to the Kerfuffle server, select the species to investigate, and click "Analyze." While the analysis is performed, the program displays a "Summary" tab which reports, for each species, the amount of genes found from the specified list, as well as invalid or missing gene names. After the completion of the analysis for each species, a tab is generated next to the summary tab that reveals a report of discovered clusters. This report includes a plot of the distribution of distances between consecutive gene pairs (Figure 2A), a histogram of cluster sizes (Figure 2B), and a genomic visualization of the clusters (Figure 3). The graphs and data points used to plot them are downloadable. At the top of each species tab, the genes in a cluster is presented along with a clickable link which searches the KEGG pathway database [22] for commonality in the cluster. Further down the tab, the interactive analysis plots are generated: clicking on a data point for a given consecutive distance will display all the consecutive gene pairs separated by that distance, as well as display the histogram counts over the graph. To assess the significance of the clustering, we overlay a plot of the expected distribution under random gene shuffling, i.e. if gene colocalization were random. Deviation from the null distribution is also quantified as a p-value table generated using a permutation test, as previously discussed. Note that the null distribution curve in the "Consecutive gene pairs" graph (Figure 2A) may sometimes appear to be linear, as opposed to the expected exponential, due to the significance of gene clustering—significant counts may "squish" the null-curve. As a result, an option is available to generate an independent plot of this curve, demonstrating the decaying nature of the distribution.

Under the table of clustered genes in the species (currently only human, mouse, rat, drosophila) tab, a Circos image may be generated for visualization of gene clusters. These images may be downloaded and saved as either .png or .svg formats. Figure 3 shows a Circos plot of the clustered genes from our online example of synapse genes. The sizes of the clusters are represented by a green histogram located at the appropriate genomic start and stop of the clustered genes, pointing radially inwards. We have attempted to optimize output for visualization of gene names (pointing radially outwards) while maintaining all genes on the image, however, some genes may run-off the Circos image because it is impossible to know *a priori* how many genes will sit next to each other in any given colocalization analysis.

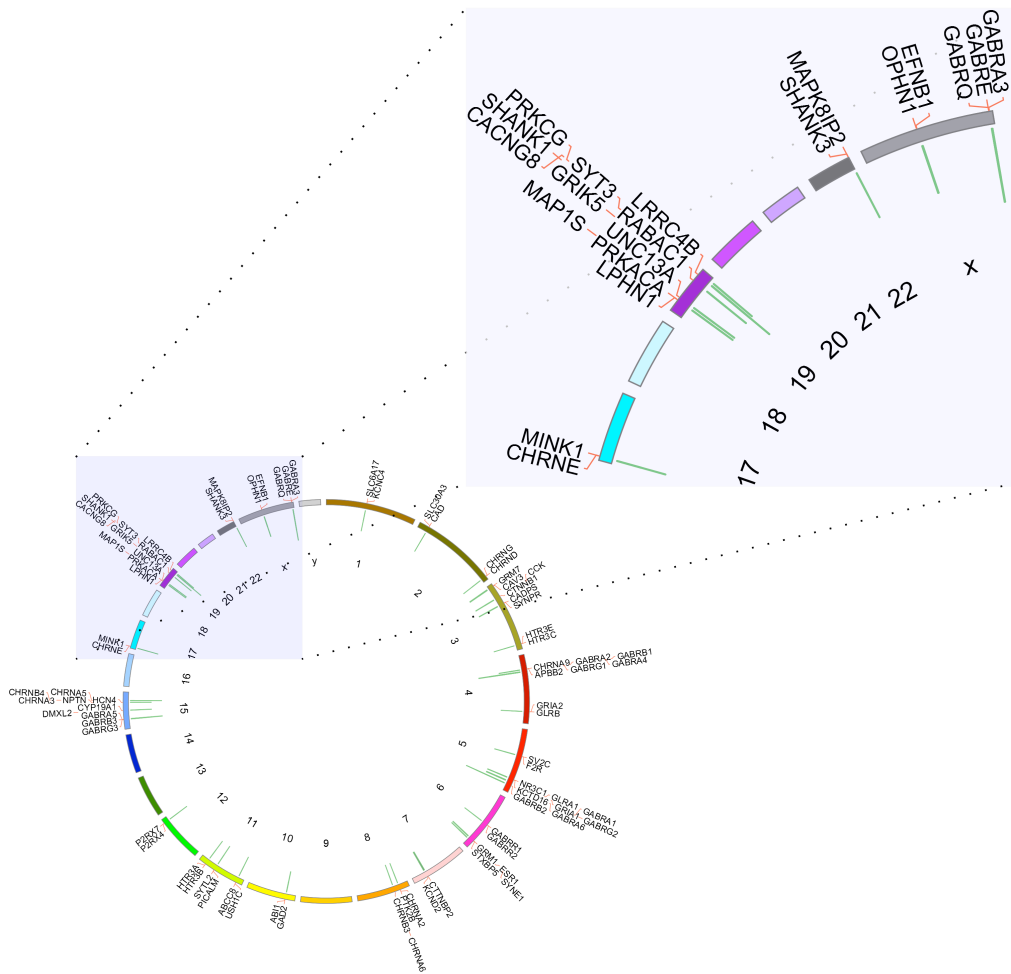

**Figure 3:** Circos output from the example clustered synapse genes. Kerfuffle outputs a Circos plot clustering the genes investigated by the user. The clusters are quantified by the green bars protruding inward in the Circos plot. The longer the bars, the more genes in the cluster. The output image also lists the colocalized genes. Scalable vector images (svg) and png images are available online

Finally, Kerfuffle makes it easy to compare cluster conservation across all species. The user may click on the "Summary" tab and run a comparative analysis by clicking "Go." Under the newly generated tab "Compare", the *Score(S,T)* statistic is calculated and displayed demonstrating the degree of conservation of the clusters in each species relative to humans. Below these results, the consecutive distance distribution for each species is simultaneously displayed (Figure 2C).

Once the user becomes familiar with the performed analyses, default parameters (discussed previously) may be changed at the bottom of the website below the "Analyze" button. Further, the uploaded genes may be dynamically removed by clicking on the "x" next to each gene or added to the list, whereby the analysis will need to be re-run to reflect the changes. The gene list may also be reset without disturbing the analyses performed and the whole set of analyses may be reset using the "New" button at the top of the page. If the user has any difficulties, we have created a comprehensive Frequently Asked Questions (FAQ) section which covers the capabilities of the whole website and answers many of the more common questions. Finally once a user account is created, all results can be saved online for later analysis.

**Performance**

To evaluate the performance of our web tool, we ran several queries using gene sets of varying size and different number of species (Figure 4). We find that a typical query of ~500 genes in 5 species completes in ~25 seconds (or ~3 minutes when querying all 47 species). Overall, for a given number of species, the running time increases exponentially with the number of input genes (Figure 4). However, even a query of 5,000 genes (an unusually high number of genes) in all 47 species completes in less than 10 minutes. Hence, our server is well suited to ensure that queries are handled expediently. Although there is no limit on how many genes a user can input, we recommend that users do not exceed 10,000 genes in order to maintain a reasonable running time, as well as the usefulness of results (too many genes increases the likelihood of finding clusters).

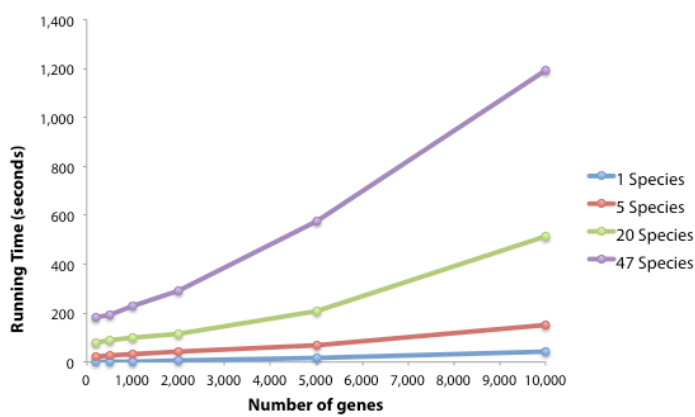

**Figure 4:** Running time of Kerfuffle as a function of the number of genes queried. For any given number of species, the running time increases exponentially with the number of input genes, but does not exceed 20 minutes for up to 10,000 genes in 47 species.

**Future Developments**

Future developments will include increased investigative options, such as changing the type of genes investigated (currently set to protein-coding only) and incorporation of other gene name schemes (such as RefSeq

IDs). Currently, our default conservation score sets humans as the relative species, i.e. for all calculations, $S$ = Homo sapiens. While our implementation is human-centric, the end-user may wish to use another relative species. In future implementations, it is expected this option will be added. Other features, such as the identification of common clusters in the species will be added, while other functionality will be included to improve our pathway investigations. Currently, we link to the KEGG website, a multi-gene pathway search. In later developments, our webpage will determine the similar pathways and display them along with the clusters. Finally, Circos uses specific karyotype files, which define the genome of the species investigated. Our current Circos implementation, however, uses the default available files: humans, mouse, rat, and drosophila. In future developments, we will generate karyotype files for any available genome, making visualization of clusters available for a much wider range of species.

## Conclusions

This software is intended for the end-user to quickly and efficiently obtain genomic organizational information about a set of user-defined functionally related genes. The software discovers clusters in each species selected and determines the significance of those clusters while allowing for interactive and visual exploration of genomic structure.

Since it is expected that speciation would lead to differences in genomic organization, provided organization is random, we investigate relative cluster conservation between species using a measure we define as the *Score(S,T)*. Once the analysis is performed, the user may compare species and determine the degree of cluster conservation.

The optional parameters make the investigations customizable and allow the user to optimize run-time. An account may also be created where all investigations may be saved for later use. Further, our website has an extensive FAQ section which may help guide the user.

## Availability and Requirements

- Project name: Kerfuffle
- Project home page: http://atwallab.org/kerfuffle
- Operating system(s): Platform independent
- Programming language: PHP, JavaScript, C++
- Other requirements: Web browser (supported browsers: Firefox, Chrome, Internet Explorer or Safari)
- License: GNU GPL
- Any restrictions to use by non-academics: no license needed.

## List of abbreviations

AJAX: Asynchronous JavaScript and XML, PHP, PHP Hypertext Preprocessor, HTML: Hypertext Markup Language, BLAST: Basic Local Alignment Search Tool, KEGG: Kyoto Encyclopedia of Genes and Genomes. RefSeq: Reference Sequence.

## Author's Contributions

RA carried out the code development, implementation, and drafted the manuscript. BF helped draft the manuscript and contributed to code and research development. GS conceived of the study, participated in its design and coordination, and helped to draft the manuscript. All authors read and approved the final manuscript.

## Competing interests

The authors declare no competing interests.

## Author's Information

Centered at Cold Spring Harbor Laboratory, RA is a graduate student in the Watson School of Biological Sciences, BF is a postdoc in the Atwal Lab, and GA is the Atwal lab head.

## Funding

This work was supported by the Starr Fellowship and the CSHL Watson School of Biological Sciences NIH training grant [5T32GM065094 to R.A.], the Simons Foundation [to B.F. and G.S.A] and the Starr Cancer Consortium [I3-A123 to G.S.A.].s

## Acknowledgements

We gratefully thank Peter Andrews for technical assistance with the web server and Ying Cai for useful comments on the web tool.

## References

1. Hurst LD, Pal C, Lercher MJ: **The evolutionary dynamics of eukaryotic gene order**. *Nat Rev Genet* 2004, **5**(4):299-310.
2. Petkov PM, Graber JH, Churchill GA, DiPetrillo K, King BL, Paigen K: **Evidence of a Large-Scale Functional Organization of Mammalian Chromosomes**. *PLoS Genet* 2005, **1**(3):e33.
3. Xue W, Kitzing T, Roessler S, Zuber J, Krasnitz A, Schultz N, Revill K, Weissmueller S, Rappaport AR, Simon J *et al*: **A cluster of cooperating tumor-suppressor gene candidates in chromosomal deletions**. *Proceedings of the National Academy of Sciences* 2012.
4. Blumenthal T: **Operons in eukaryotes**. *Briefings in functional genomics & proteomics* 2004, **3**(3):199-211.
5. Lee JM, Sonnhammer ELL: **Genomic Gene Clustering Analysis of Pathways in Eukaryotes**. *Genome Research* 2003, **13**(5):875-882.
6. Carroll SB: **Homeotic genes and the evolution of arthropods and chordates**. *Nature* 1995, **376**(6540):479-485.
7. Lercher MJ, Urrutia AO, Hurst LD: **Clustering of housekeeping genes provides a unified model of gene order in the human genome**. *Nat Genet* 2002, **31**(2):180-183.
8. Singer GAC, Lloyd AT, Huminiecki LB, Wolfe KH: **Clusters of Co-expressed Genes in Mammalian Genomes Are Conserved by Natural Selection**. *Molecular Biology and Evolution* 2005, **22**(3):767-775.
9. Al-Shahrour F, Minguez P, Marqués-Bonet T, Gazave E, Navarro A, Dopazo J: **Selection upon Genome Architecture: Conservation of Functional Neighborhoods with Changing Genes**. *PLoS Comput Biol* 2010, **6**(10):e1000953.
10. Yi G, Sze SH, Thon MR: **Identifying clusters of functionally related genes in genomes**. *Bioinformatics* 2007, **23**(9):1053-1060.
11. Revanna KV, Krishnakumar V, Dong Q: **A web-based software system for dynamic gene cluster comparison across multiple genomes**. *Bioinformatics* 2009, **25**(7):956-957.
12. Vilella AJ, Severin J, Ureta-Vidal A, Heng L, Durbin R, Birney E: **EnsemblCompara GeneTrees: Complete, duplication-aware phylogenetic trees in vertebrates**. *Genome Research* 2009, **19**(2):327-335.

13. Proost S, Fostier J, De Witte D, Dhoedt B, Demeester P, Van de Peer Y, Vandepoele K: **i-ADHoRe 3.0—fast and sensitive detection of genomic homology in extremely large data sets**. *Nucleic acids research* 2012, **40**(2):e11-e11.
14. Wang Y, Tang H, DeBarry JD, Tan X, Li J, Wang X, Lee T, Jin H, Marler B, Guo H: **MCScanX: a toolkit for detection and evolutionary analysis of gene synteny and collinearity**. *Nucleic acids research* 2012, **40**(7):e49-e49.
15. Sinha A, Meller J: **Cinteny: flexible analysis and visualization of synteny and genome rearrangements in multiple organisms**. *BMC Bioinformatics* 2007, **8**(1):82.
16. Ng M-P, Vergara I, Frech C, Chen Q, Zeng X, Pei J, Chen N: **OrthoClusterDB: an online platform for synteny blocks**. *BMC Bioinformatics* 2009, **10**(1):192.
17. Rödelsperger C, Dieterich C: **Syntenator: multiple gene order alignments with a gene-specific scoring function**. *Algorithms Mol Biol* 2008, **3**:14.
18. Kinsella RJ, Kähäri A, Haider S, Zamora J, Proctor G, Spudich G, Almeida-King J, Staines D, Derwent P, Kerhornou A *et al*: **Ensembl BioMarts: a hub for data retrieval across taxonomic space**. *Database* 2011, **2011**.
19. Krzywinski M, Schein J, Birol İ, Connors J, Gascoyne R, Horsman D, Jones SJ, Marra MA: **Circos: an information aesthetic for comparative genomics**. *Genome Research* 2009, **19**(9):1639-1645.
20. Carbon S, Ireland A, Mungall CJ, Shu SQ, Marshall B, Lewis S: **AmiGO: online access to ontology and annotation data**. *Bioinformatics* 2009, **25**(2):288-289.
21. Paul A, Cai Y, Atwal GS, Huang ZJ: **Developmental coordination of gene expression between synaptic partners during GABAergic circuit assembly in cerebellar cortex**. *Frontiers in Neural Circuits* 2012, **6**.
22. Kanehisa M, Goto S, Sato Y, Furumichi M, Tanabe M: **KEGG for integration and interpretation of large-scale molecular data sets**. *Nucleic acids research* 2012, **40**(D1):D109-D114.